# Decorated electronic kagome lattice in twisted bilayer germanene

Dennis J. Klaassen[1&], Rian A.M. Ligthart[2], Andrés R. Botello Mendez[2], Lumen Eek[3], Esra van 't Westende[1], Paul L. de Boeij[1], Y. Wang[1], D. Vanmaekelbergh[2], Pantelis Bampoulis[1], Cristiane Morais Smith[3], Ingmar Swart[2], Zeila Zanolli[2], and Harold J.W. Zandvliet[1&]

[1] Physics of Interfaces and Nanomaterials, MESA+ Institute, University of Twente, P.O. Box 217, 7500AE Enschede, The Netherlands

[2] Debye Institute for Nanomaterials Science, Utrecht University, P.O. Box 80.000, 3508TA Utrecht, The Netherlands

[3] Institute for Theoretical Physics, Utrecht University, Princetonplein 5, 3584CC Utrecht, The Netherlands

Artificial kagome lattices provide a route to electronic flat bands, geometric frustration, and correlation driven phases, but their realization in atomically controlled two-dimensional materials remains scarce. Here, we show that commensurate twisted bilayer germanene on $Ge_2Pt$ produces two electronically distinct large-angle moiré phases. Scanning tunneling microscopy measurements and density functional theory calculations reveal that commensurate twisted bilayers that are odd under an exchange of sublattices are semiconducting, whereas the twisted bilayers with an even parity are metallic. The twisted bilayers with an even parity host an empty state resonance that exhibits an emergent decorated kagome structure with a $C_3$ symmetry. These results establish large-angle twisted germanene as a platform for engineering kagome-like electronic states in a buckled two-dimensional material.

&) corresponding authors: d.j.klaassen@utwente.nl and h.j.w.zandvliet@utwente.nl

## 1. Introduction

Kagome lattices are a central motif in quantum materials because their corner-sharing triangular geometry can generate geometric frustration, Dirac-like dispersions, van Hove singularities and flat energy bands [1-4]. Kagome lattices play a pivotal role in the study of frustrated magnetism and spin liquids. The key to the emergence of frustrated magnetism in a kagome lattice is that an antiferromagnetic Heisenberg model on a triangular lattice has no unique ground state that satisfy the antiferromagnetic correlation interactions. The latter results in spin frustration and strong quantum fluctuations. Apart from the relevance for magnetism, kagome lattices also exhibit an intriguing nontrivial topological electronic structure [3].

Materials exhibiting a kagome structure, however, are relatively uncommon, motivating efforts to create artificial kagome lattices in tunable two-dimensional platforms. Twisted van der Waals materials offer a powerful route to such artificial electronic lattices [5-8]. In twisted bilayer graphene, emergent kagome electronic patterns have been predicted and observed under special conditions, most prominently at very small twist angles or near selected commensurate configurations [5-8]. In these systems, moiré reconstruction, strain, and interlayer coupling can reorganize the local density of states into kagome-like real-space patterns. Graphene, however, is a planar honeycomb material with weak intrinsic spin orbit coupling. Buckled honeycomb materials, like silicene and germanene, could provide a new route for realizing kagome phases combined with topological properties. In germanene, the two sublattices are vertically displaced and the spin-orbit interaction is much stronger than in graphene and silicene [9-16]. As a result, interlayer registry, sublattice asymmetry, and local electric fields are expected to be coupled directly to the electronic structure. This makes twisted bilayer germanene a natural platform for investigating how kagome-like electronic states form in a buckled honeycomb lattice with spin-orbit coupling [17-19]. In such a system, the twist angle does not only lead to a moiré superstructure but also organizes locally distinct stacking configurations between upward- and downward-buckled atoms. These registries can lower the symmetry of the emergent electronic lattice and may produce novel kagome patterns that have no direct analogue in planar graphene.

Here we show that commensurate twisted bilayer germanene on $Ge_2Pt$ realizes a kagome pattern. Scanning tunneling microscopy and density functional theory calculations reveal two dominant large-angle phases: a semiconducting $(\sqrt{7} \times \sqrt{7})$ phase that is odd under sublattice exchange and a metallic $(\sqrt{13} \times \sqrt{13})$ phase that is even under sublattice exchange. The metallic phase exhibits a sharp empty-state resonance at 1.45 eV. Spatial mapping of this resonance reveals an emergent decorated kagome electronic lattice with $C_3$ symmetry. A minimal tight-binding model shows that the decoration breaks inversion and $C_6$ symmetry, while preserving the characteristic kagome flat band (see SI).

## 2. Methods

### 2.1 Experimental setup and sample preparation

The experiments were performed in two different ultra-high vacuum systems with base pressures below $2 \times 10^{-10}$ mbar and $3 \times 10^{-11}$ mbar, respectively. The systems are equipped with a Unisoku USM1300 and an Omicron low-temperature scanning tunneling microscope. We have used intrinsic Ge(110) substrates, which were ultrasonically cleaned in isopropyl alcohol before inserting into the vacuum system. In the ultra-high vacuum system, the Ge(110) substrates were degassed for 24 hours at a temperature of 700 K. Subsequently, the Ge(110) substrates were cleaned by several cycles of 800 eV $Ar^+$ bombardment and annealing at 1100 K via resistive heating. After cleaning, a few monolayers of Pt were deposited onto the Ge(110) substrate at room temperature followed by a 2 minute anneal at a temperature of 1100 K. This procedure leads, as described in detail in our previous papers [10,16], to the formation of a eutectic $Ge_{0.78}Pt_{0.22}$ alloy. Upon slowly cooling, the eutectic alloy undergoes spinodal decomposition, resulting in the formation of $Ge_2Pt$ nanocrystals with typical dimensions in the range of a few hundreds of nm to a few μm, which are coated with a few germanene layers. Some of these germanene layers are twisted with respect to each other, resulting in moiré superlattices. The scanning tunneling microscopy and spectroscopy measurements were performed at 77 K, unless stated otherwise.

### 2.2 Density functional theory calculations

To complement the experimental observations, we performed first-principles density functional theory (DFT) calculations. The electronic structures were modeled within the localized pseudo-atomic orbital framework, as implemented in the SIESTA package [20]. To accurately capture the properties of buckled germanene, spin-orbit coupling (SOC) was included in all calculations. The exchange-correlation functional was treated within the generalized gradient approximation (GGA) in the Perdew-Burke-Ernzerhof (PBE) formulation. The SIESTA calculations were benchmarked against plane-wave calculations performed using the Quantum ESPRESSO (QE) package [21], with norm-conserving pseudopotentials and a plane-wave kinetic energy cutoff of 95 Ry. The benchmarking showed excellent agreement: the positions of the flat bands relative to the Fermi level ($E_F$) agreed within 0.02–0.03 eV, and their bandwidths matched within 5%, validating the localized orbital approach. For structural optimization and lattice constant determination, Quantum ESPRESSO variable-cell relaxations were performed using the PBE, PBEsol exchange-correlation functionals [22], as well as two van der Waals (vdW) correction schemes: a non-local vdW density functional (vdW-DF) [23] and the semi-empirical Grimme D3 correction [24], finding similar results, but qualitatively different from non-relaxed structures. For the SIESTA calculations, we used a dense $24 \times 24 \times 1$ Monkhorst-Pack $k$-point mesh and a mesh cutoff of 600 Ry, yielding a highly converged description of the moiré Brillouin zone. The resulting tight-binding Hamiltonians and wavefunctions were post-processed using the *sisl* library [25] for layer projection and bonding analysis.

## 3. Results and Discussion

In Figures 1(a) and 1(b), we show two scanning tunneling microscopy images of twisted bilayer germanene. The moiré patterns in Figures 1(a) and 1(b) have periodicities of 1.1 nm and 1.5 nm, corresponding to the commensurate $\sqrt{7} \times \sqrt{7}$ and $\sqrt{13} \times \sqrt{13}$ structures, respectively. In our experiments these structures are the most abundant moiré structures. In Figures 1(c) and 1(d) the corresponding scanning tunneling spectra are displayed. These differential conductivity spectra reveal that the $\sqrt{7} \times \sqrt{7}$ structure is semiconducting with a sizeable band gap of ~ 0.2 eV, whereas the $\sqrt{13} \times \sqrt{13}$ structure is metallic.

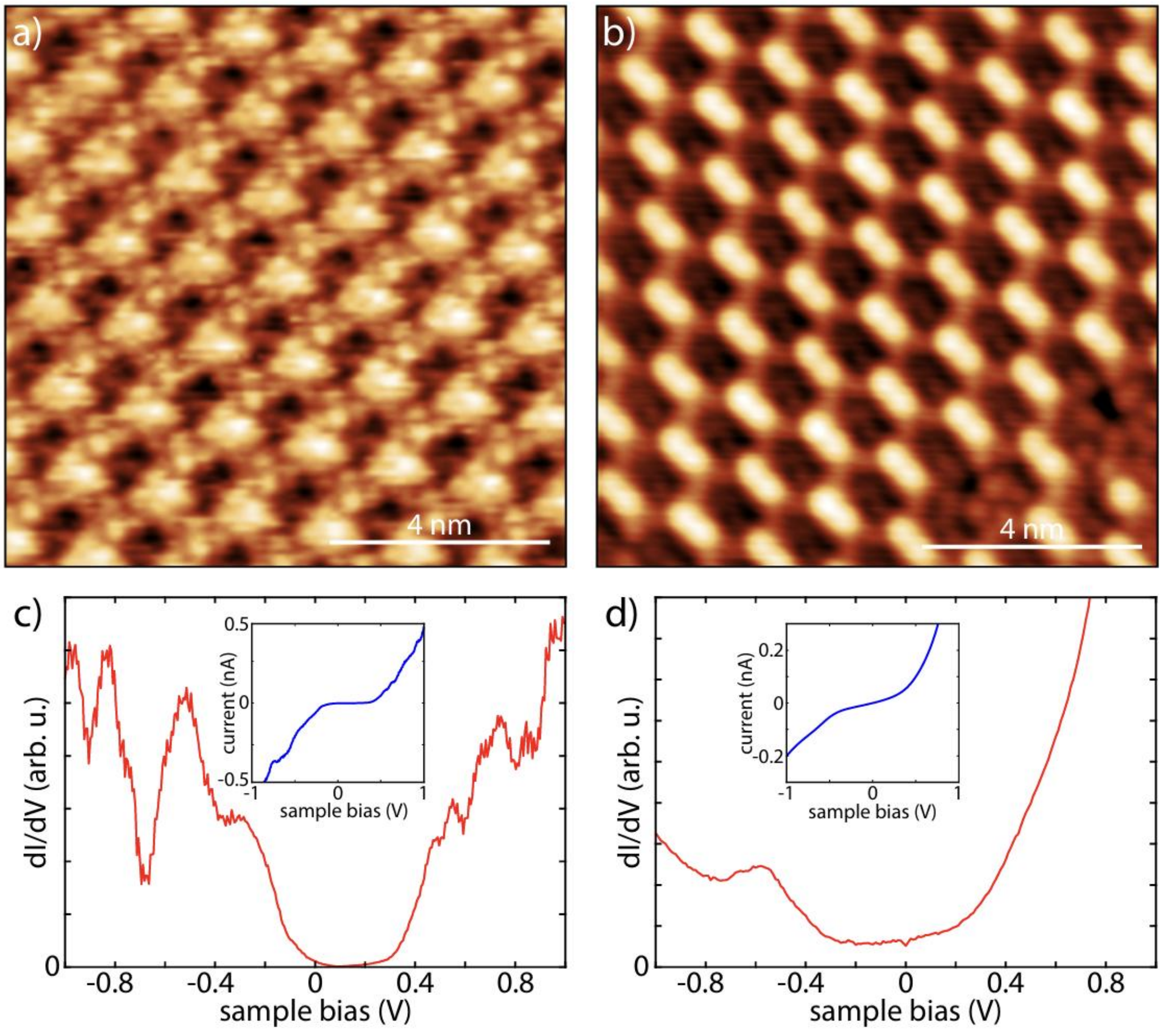

***Figure 1*** *Scanning tunneling microscopy images and differential conductivity of twisted bilayer germanene. (a) Scanning tunneling microscopy image of* $\sqrt{7}\times\sqrt{7}$ *twisted bilayer germanene, sample bias -1.3 V and tunnel current 0.3 nA (b) Scanning tunneling microscopy image of* $\sqrt{13}\times\sqrt{13}$ *twisted bilayer germanene, sample bias -0.05 V and tunnel current 0.5 nA, T= 4.2 K (c) Differential conductivity (dI(V)/dV) of* $\sqrt{7}\times\sqrt{7}$ *twisted bilayer germanene. The inset shows the I(V) curve. Setpoint sample bias –1.5 V and setpoint tunnel current 1 nA. and (d) Differential conductivity (dI(V)/dV) of* $\sqrt{13}\times\sqrt{13}$ *twisted bilayer germanene. The inset shows the I(V) curve, T=0.36 K. Setpoint sample bias -1 V and setpoint tunnel current 0.2 nA.*

### 3.1 Structural properties of twisted bilayer germanene

Before discussing the electronic properties of the twisted bilayers, we will first elaborate on the structural properties of the twisted germanene bilayers. We define the twist angle with respect to the AA stacking configuration and use one of the atoms of the top layer as a rotation point. Like commensurate twisted bilayer graphene, there are two different commensurate twisted bilayer germanene configurations: structures that are odd and structures that are even under an exchange of sublattices [26]. The twist angle of the odd $\sqrt{7} \times \sqrt{7}$ structure is $21.8^o$, whereas the twist angle of the even $\sqrt{7} \times \sqrt{7}$ structure is $38.2^o$. Likewise, we find for the $\sqrt{13} \times \sqrt{13}$ structure twist angles $32.2^o$ (odd) and $27.8^o$ (even), respectively. Based on an analysis of the scanning tunneling microscopy image and its Fourier transform, we find that the realized $\sqrt{7} \times \sqrt{7}$ structure has a twist angle of $21.8^o$, corresponding to the odd symmetry, see ref. [19] for further details. The $\sqrt{13} \times \sqrt{13}$ structure, is, as we will show below, even under an exchange of sublattices and has a twist angle of $27.8^o$.

In Figures 2(a)-(b) we show ball-and-stick models of the odd $\sqrt{7} \times \sqrt{7}$ and even $\sqrt{13} \times \sqrt{13}$ structures, respectively. In the odd structure, there is one A atom per moiré unit cell of the top layer that sits exactly atop of an A atom of the bottom layer and there are two B atoms (one from the top layer (dark red) and one from the bottom layer (dark blue)) that are in the center of a hexagon, see Figure 2(a). Following the nomenclature used in the literature [26], we refer to the odd structure as an AB-like structure. In the even structure, one finds per moiré unit cell one A atom and one B atom of the top layer located on an A atom and a B atom of the bottom layer. In addition, the even structure also has a region where the hexagons of the top and bottom layer eclipse, see Figure 2(b). The even structure has some similarities with an AA stacked bilayer and is therefore referred to as AA-like.

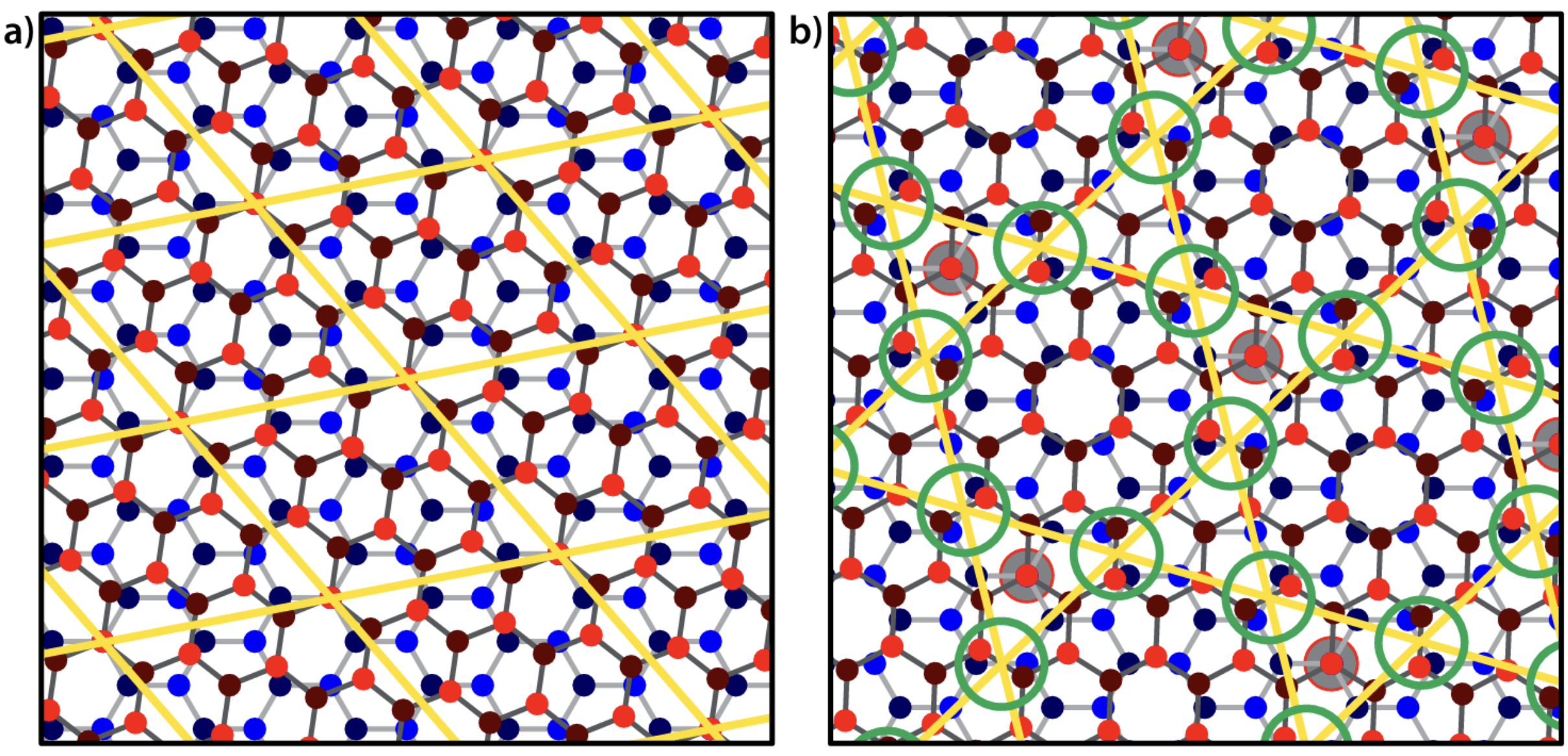

***Figure* 2** *Schematic models of the odd* $(\sqrt{7}\times\sqrt{7})$ *and even* $(\sqrt{13}\times\sqrt{13})$ *unit cells of a twisted bilayer germanene. The light (dark) red dots represent upward (downward) buckled atoms of the top layer. The light (dark) blue dots represent upward (downward) buckled atoms of the bottom layer. The large green open circles in (b), which are exactly in the middle between two adjacent stacked AA regions, form an emergent kagome lattice. The kagome structure is outlined by the yellow lines. The corner-sharing triangles either have an upward buckled atom (light red) or a downward buckled atom (dark red) in their center. The upward buckled atoms in the centers of the corner-sharing triangles are highlighted by a grey circle.*

### 3.2 Comparison between theory and experiments

Next, we will compare our scanning tunneling spectroscopy results of the odd and even twisted bilayers germanene with ab-initio calculations. As shown in Figures 1(c)-(d), the odd $\sqrt{7}\times\sqrt{7}$ structure is semiconducting with a band gap of ~ 0.2 eV, and the even $\sqrt{13}\times\sqrt{13}$ structure is metallic. This salient electronic difference is also found in our density functional theory calculations and can be explained by a combination of symmetry and chemical bonding arguments. In buckled twisted bilayer germanene, the out-of-plane buckling height breaks the horizontal mirror plane, giving rise to configurations that are either odd (AB-like) or even (AA-like) under the exchange of sublattices. For the odd $\sqrt{7}\times\sqrt{7}$ structure, the stacking registry breaks inversion symmetry and layer degeneracy. This broken layer symmetry is reflected in the wavefunctions of states near $E_F$, which exhibit highly asymmetric top/bottom layer weights, as resolved by our bonding analysis (crystal orbital overlap population and crystal orbital Hamilton population). This layer asymmetry splits the degenerate bonding and antibonding states, pushing the flat bands ~ 0.26 eV above the Fermi level and opening a clean semiconducting gap. In Figure 3(a), we show the calculated electronic band structure of the odd $\sqrt{7} \times \sqrt{7}$ AB-like structure. There is a clear energy gap at $E_F$, matching the experimental scanning tunneling spectrum in Figure 1(c). Chemically, the AB registry positions atoms of opposite sublattices at different buckling heights directly across the van der Waals gap, creating highly effective interlayer orbital overlap. Mayer bond order analysis reveals that the AB stacking has 15% more interlayer bonding (Mayer bond order of 17.2 vs. 14.9 in the AA stacking) and 13% stronger average interlayer hopping ($\langle|H_{\text{inter}}|\rangle = 0.093$ eV vs. 0.082 eV in AA). This enhanced interlayer hybridization stabilizes the bonding states below $E_F$ while destabilizing the antibonding states above $E_F$, shifting the flat band and opening the gap. In contrast, the even $\sqrt{13} \times \sqrt{13}$ AA-like structure preserves layer exchange symmetry (even parity). This guarantees that every state near the Fermi level has exactly equal weight on both layers (perfect 50/50 top/bottom projection). This symmetry enforces degenerate bonding/antibonding pairs, preventing the opening of a gap and keeping the flat bands at the Fermi level, rendering the even systems metallic. As a final remark, we would like to refer to related work on twisted bilayer graphene [26, 27]. Mele pointed out that even though the odd and even structures of commensurate twisted bilayer graphene are very similar to AB and AA stacked bilayer graphene, the energy scales of their low-energy bands are substantially inflated. This inflated energy scale allows tuning of the band gap in the odd structure by relatively small electric fields, whereas the even structure is virtually immune to an applied electric field [27]. The difference in work function between PtIr scanning tunneling microscopy tip and germanene substrate results in a built-in electric field in the scanning tunneling microscopy junction [28,29]. Similar to twisted bilayer graphene, this electric field is expected to affect the band gap of odd, but not even bilayers germanene.

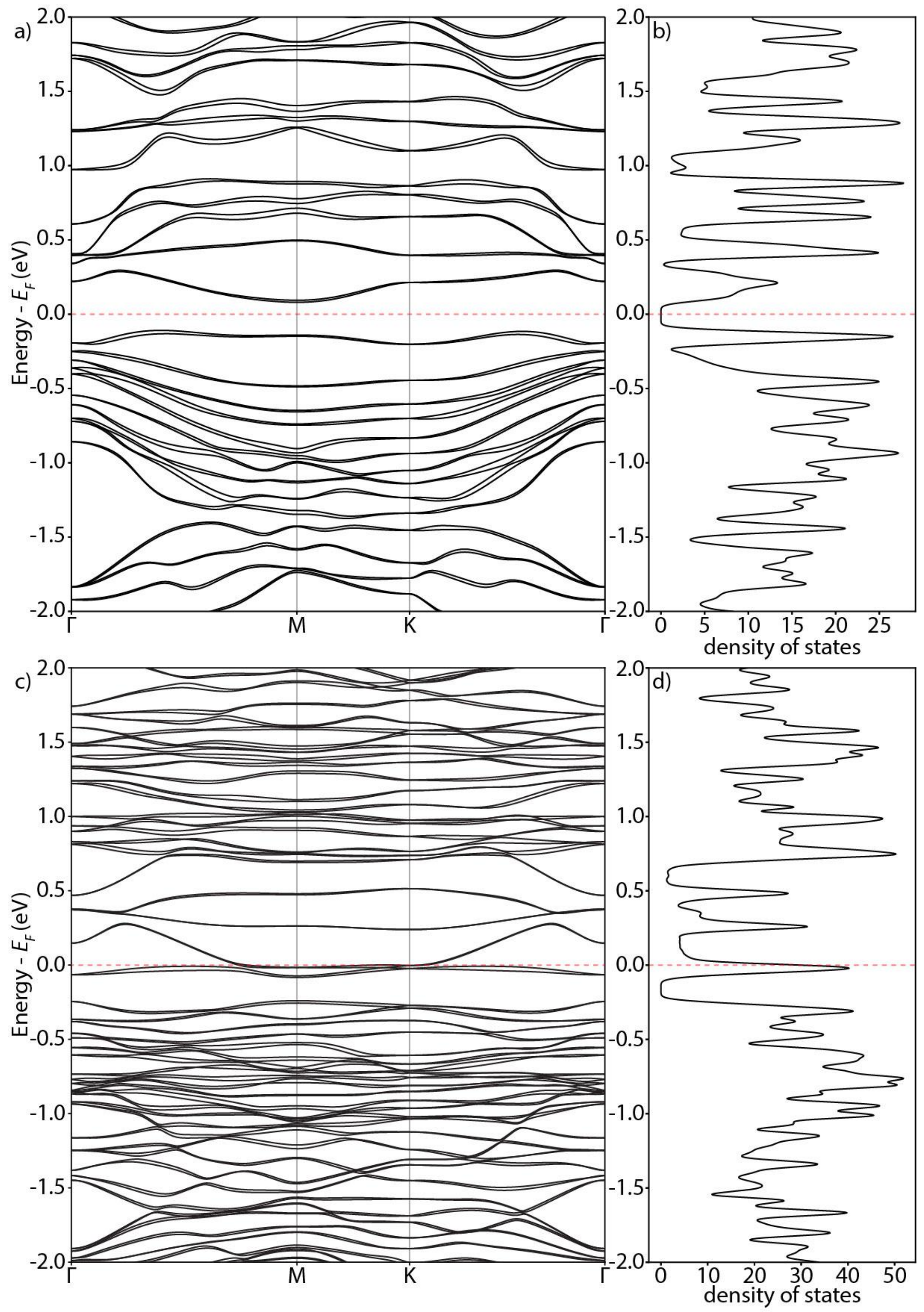

***Figure 3*** *(a) Density Functional Theory (DFT) calculations of the odd* $\sqrt{7} \times \sqrt{7}$ *twisted bilayer germanene structure. Electronic band structure along high-symmetry paths in the moiré Brillouin zone, showing flat bands near the Fermi level, and at high energy, and (b) Corresponding density of states showing prominent peaks at the flat-band energies. (c) Density Functional Theory (DFT) calculations of the even* $\sqrt{13} \times \sqrt{13}$ *twisted bilayer germanene structure. Electronic band structure along high-symmetry paths in the moiré Brillouin zone, showing flat bands near the Fermi level, and at high energy, and (d) Corresponding density of states showing prominent peaks at the flat-band energies.*

Next, we will focus our attention on the high-energy (> 0.5 eV) electronic band structure of the twisted $(\sqrt{13} \times \sqrt{13})$ bilayer germanene. In Figure 4(a), we show the differential conductivity at positive sample bias of the even $\sqrt{13} \times \sqrt{13}$ structure, which exhibits a few small peaks at 0.5 eV, 0.7 eV, 1.0 eV, and two substantial stronger peaks at 1.45 eV and 1.9 eV, respectively. In Figure 4(b), we show a spatial map of the 1.45 eV peak. This spatial map reveals a decorated kagome lattice (see inset). Interestingly, a very comparable scanning tunneling spectrum has been reported for twisted bilayer silicene, albeit with a small shift in energy [17,18]. Like twisted bilayer germanene, a spatial map of the peak at 1.3 V of 21.8° twisted bilayer silicene on Ag(111) also reveals an emergent electronic kagome lattice [18]. The twisted bilayer silicene displays, similar twisted bilayer germanene, a second strong peak at an energy of 1.7 eV. This second peak does not, also like twisted bilayer germanene, exhibit a kagome structure.

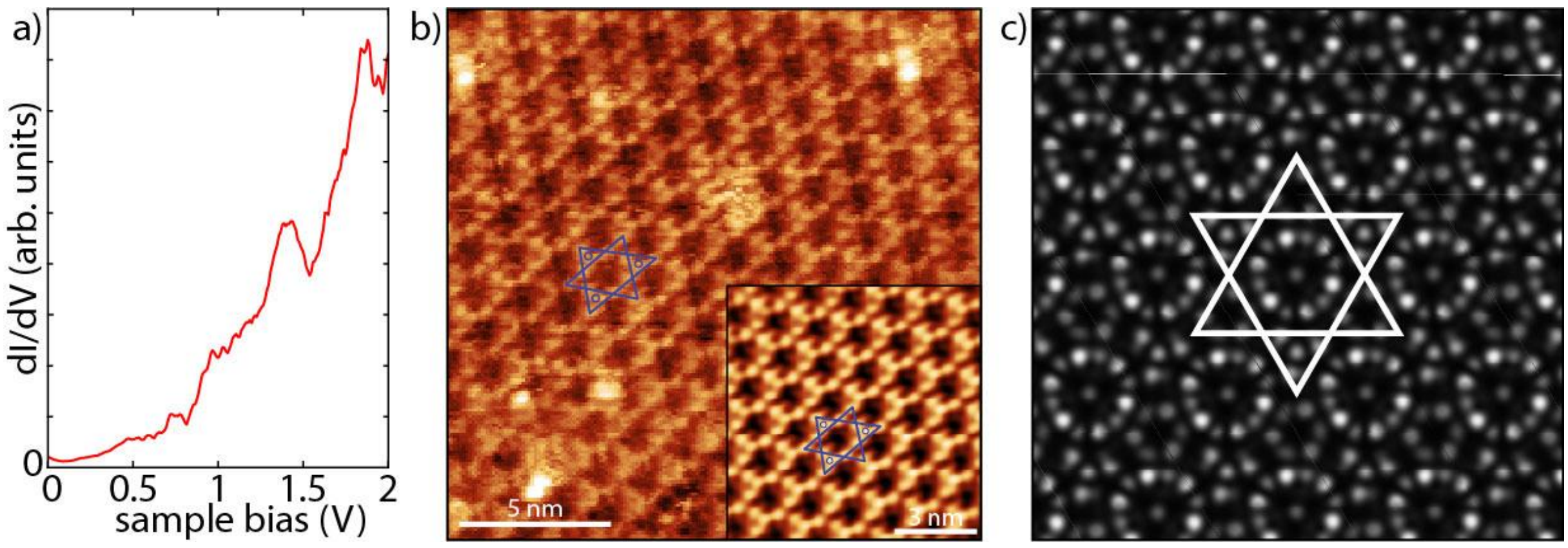

***Figure 4*** *(a) Differential conductivity of the empty states of an even* $(\sqrt{13} \times \sqrt{13})$ *twisted bilayes germanene. The peak at 1.45 V is the kagome flat band. Set points: 15 pA and -0.5 V, (b) Scanning tunneling microscopy image of an even* $(\sqrt{13} \times \sqrt{13})$ *structure taken at 1.45 V. Tunnel current is 0.1 nA. The image reveals and emergent decorated electronic kagome lattice (see inset for a filtered image). (c) Local density of states of even* $(\sqrt{13} \times \sqrt{13})$ *twisted bilayer germanene. The decorated kagome lattice is outlined by white lines.*

It is important to point out here that the silicene layers on Ag(111) do not have a regular honeycomb lattice. The silicene layers have a $(\sqrt{3} \times \sqrt{3})$ unit cell with a lattice constant of 0.64 nm [18]. In this $(\sqrt{3} \times \sqrt{3})$ unit cell only one Si atom out of the six Si atoms is buckled upwards. Twisting two layers of $(\sqrt{3} \times \sqrt{3})$ reconstructed silicene by 21.8° results in a $(\sqrt{21} \times \sqrt{21})$ moiré unit cell with a periodicity of 1.7 nm.

The even-parity $(\sqrt{13} \times \sqrt{13})$ structure exhibits quasi-flat bands near the Fermi level, arising from the moiré superlattice potential and the buckled geometry, see

Figure 3(c)-(d). A flat band is resolved at $\sim$ 1.41 eV above $E_F$, with a bandwidth of only 0.038 eV. There are also peaks in the density of states at $\sim 0.25\ eV$, $\sim 0.5\ eV$, $\sim 0.7\ eV$, and $\sim 1.0\ eV$, which, apart from the peak at $\sim 0.25\ eV$, are also found in the experimental data shown in Figure 4(a). To compare the DFT calculations to the scanning tunnelling microscopy images and verify the presence of a kagome-like decoration, we plot the local density of states with an energy window integrated from the Fermi energy to 1.5 eV. Figure 4(c) shows a two-dimensional slice parallel to the $x$-$y$ plane of the local density of states close to the surface of one of the two equivalent layers. The white lines are a guide to the eye, highlighting the decorated kagome structure. The agreement between the spatial map of the integrated local density of states and the structural model shown in Figure 2(b) is striking. The white spots near the corners and in the centre of the triangles correspond to the upward buckled Ge atoms of the top-layer (light red atoms in Figure 2(b)). A minimal tight-binding model shows that the decoration breaks inversion symmetry and $C_6$ symmetry while preserving the characteristic kagome flat band (see Appendix). The flat band at ~1.4 eV is dominated by the $d$-orbitals of the Ge atoms (66% $d$-character at $\Gamma$). The heavy atomic mass of Ge induces a large SOC splitting that separates the $p$ and $d$ contributions into distinct bands. The $d$-dominated flat band is consistent with the kagome physics, as the geometric frustration in the kagome lattice naturally produces $d$-orbital flat bands with quenched kinetic energy.

In small-angle twisted bilayer graphene, emergent kagome lattices and flat bands have been predicted and observed [8-10]. However, these graphene-based systems require tiny twist angles ($< 1°$) or large-period commensurate approximations to overcome the weak interlayer coupling of the planar lattice. In contrast, the buckled twisted bilayer germanene hosts flat bands and emergent kagome structures at large twist angles because the buckled geometry and $sp^2$-$sp^3$ hybridization produce extended orbitals with a strong out-of-plane component, ensuring that robust interlayer coupling persists even at large twist angles [19,30]. The near-zero total energy difference between the AA and AB configurations (8 meV per cell) suggests that both stackings are thermodynamically accessible, explaining the experimental co-existence of metallic and semiconducting domains.

**Conclusions**

We studied the commensurate $\sqrt{7} \times \sqrt{7}$ and $\sqrt{13} \times \sqrt{13}$ twisted bilayers germanene with scanning tunneling microscopy and density functional theory calculations. The $\sqrt{7} \times \sqrt{7}$ structure has a twist angle of 21.8°, is semiconducting and is odd under an exchange of sublattices. The $\sqrt{13} \times \sqrt{13}$ structure has a twist angle of 27.8°, is metallic and even under an exchange of sublattices. Only the metallic structure, i.e. the $\sqrt{13} \times \sqrt{13}$ structure, hosts a flat band. This flat band, which is located at 1.45 eV, exhibits an emergent decorated electronic kagome lattice with a $C_3$ symmetry. The decorated structure emerges because there are two different types of triangles in kagome structure. Half of triangles of the kagome lattice have an upward buckled atom in their center, whereas the other half of triangles have a downward buckled atom in their center.

**Acknowledgements**

L.E, D.V., I.S., Z.Z., C.M.S and H.J.W.Z., acknowledge the research program "Materials for the Quantum Age" (QuMat) for financial support. This program (registration number 024.005.006) is part of the Gravitation program financed by the Dutch Ministry of Education, Culture and Science (OCW). H. J. W. Z. and D. J. K. acknowledge funding from NWO Grant No. OCENW.M20.232. P. B. acknowledges funding from the Dutch Research Council (NWO, Grant No. OCENW.M22.123 and NWO Vidi.233.019). I.S. acknowledges funding from the European Research Council (ERC, FRACTAL, No. 865570). P.B. and Y.W. acknowledge funding from the European Research Council, funded by the European Union (ERC, Q-EDGE, No. 101162852).

# Supplementary information

## Kagome and decorated kagome lattices

The unit cell of a conventional kagome lattice has three atoms, see Figure S1(a). The electronic band structure of a kagome lattice within the framework of a simple tight-binding model that only involves nearest neighbor hopping is shown in Figure S1(b).

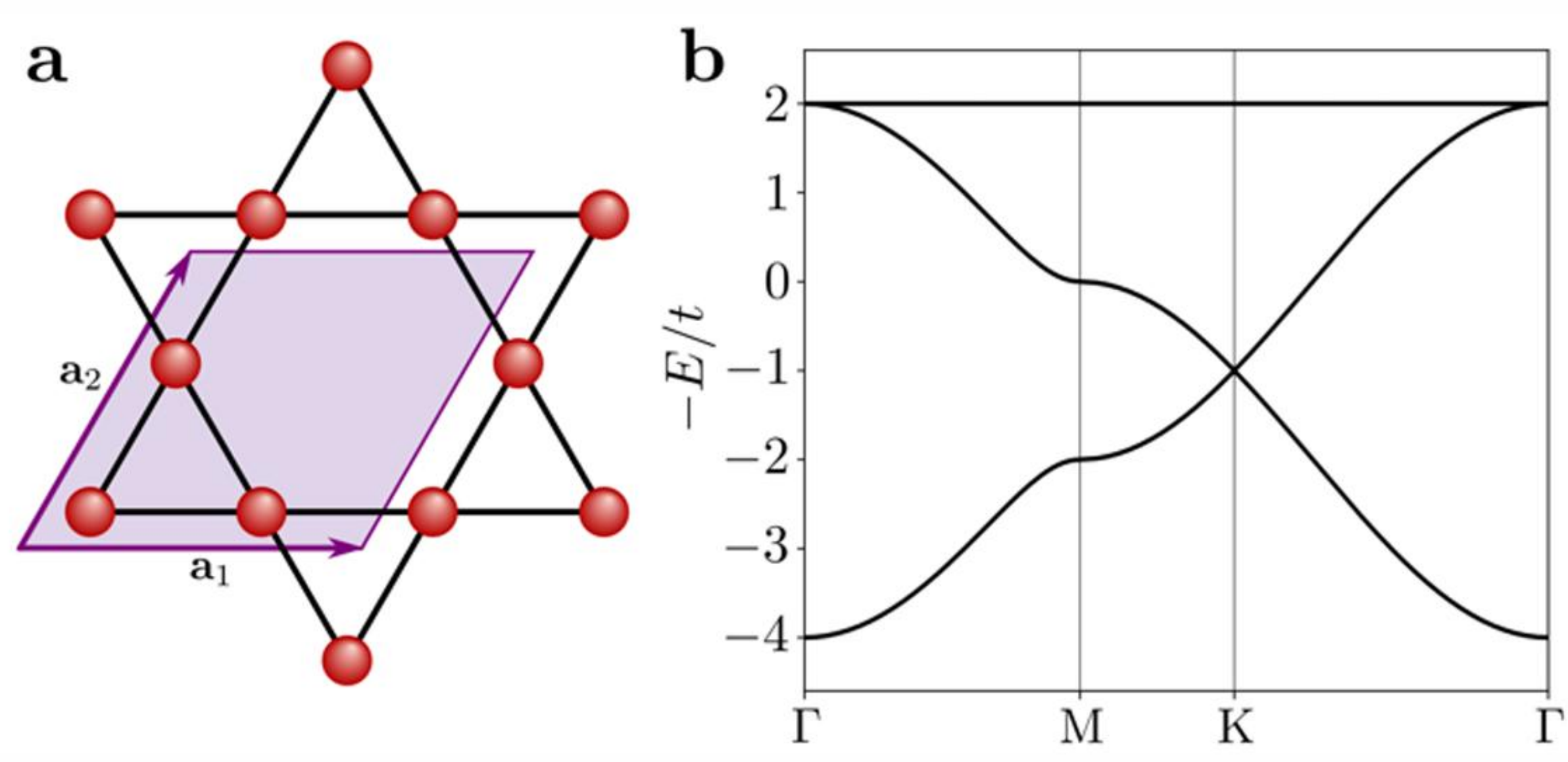

***Figure S1*** *(a) Structural model of a kagome lattice, (b) Electronic band structure of a kagome lattice.*

The kagome lattice has two linear energy bands that cross at the $K$ point of the Brillouin zone at the hopping energy $t$ ($t < 0$) and a completely dispersionless band located at an energy $-2t$, where $t$ is the nearest-neighbor hopping parameter. The electrons in this flat band are localized owing to destructive interference of the wave functions of the different sublattices [31,32]. In addition, there is a van Hove singularity near the Fermi level, which makes the kagome lattice metallic. The decorated kagome lattice is different from a conventional kagome lattice in the sense that there is an extra atom in half of the corner-sharing triangles. In Figure S2 we show the electronic band structure of the decorated kagome lattice as obtained with tight-binding calculations involving only nearest neighbor hopping.

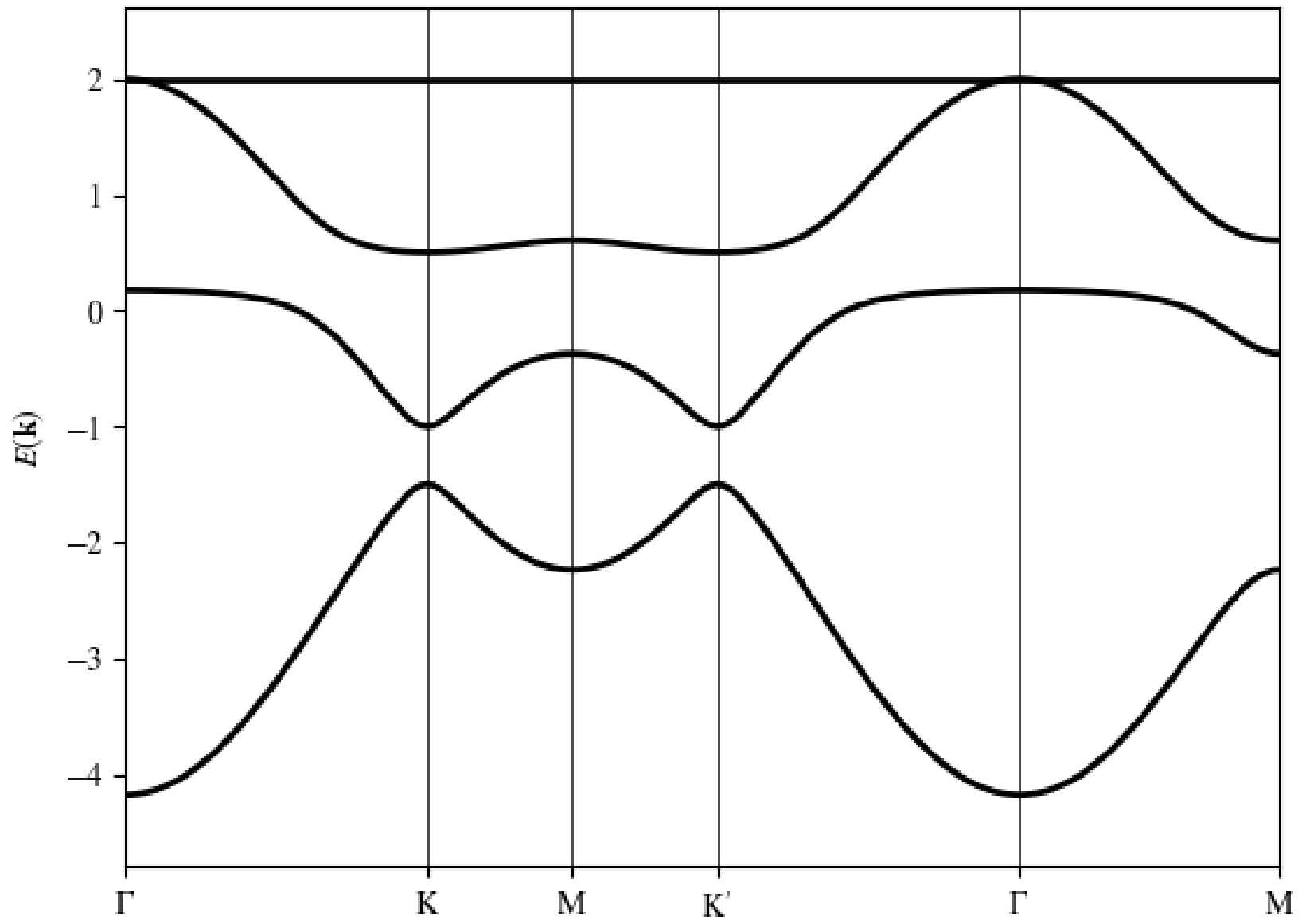

***Figure S2** Electronic band structure of a decorated kagome lattice as obtained with tight binding calculations.*

The extra atom in the unit cell of the decorated kagome lattice leads to an additional energy band. The decoration of the kagome lattice breaks inversion and $C_6$ rotational symmetry, resulting in the opening of a band gap at the Dirac points. The characteristic kagome flat band remains, however, preserved.

To scrutinize the effect of this additional fourth atom in the unit cell, we first consider a conventional kagome lattice, which is spanned by the Bravais lattice vectors $\boldsymbol{a}_1 = a\left(\frac{1}{2}\sqrt{3}, -\frac{1}{2}\right)$ and $\boldsymbol{a}_2 = a\left(\frac{1}{2}\sqrt{3}, \frac{1}{2}\right)$, see Figure S1. The three atoms in the unit cell have coordinates,

$$\boldsymbol{r}_1 = \frac{1}{2}\boldsymbol{a}_1, \quad r_2 = \frac{1}{2}\boldsymbol{a}_2, \quad \text{and} \quad \boldsymbol{r}_3 = \frac{1}{2}(\boldsymbol{a}_1 + \boldsymbol{a}_2). \tag{1a}$$

For notational convenience we define,

$$\boldsymbol{a}_3 = \boldsymbol{a}_2 - \boldsymbol{a}_1 \tag{1b}$$

The tight-binding Hamiltonian of the kagome lattice that only involves nearest-neighbor interactions is given by,

$$H(\boldsymbol{k}) = \epsilon I_3 + 2t \begin{bmatrix} 0 & c_3(\boldsymbol{k}) & c_2(\boldsymbol{k}) \\ c_3(\boldsymbol{k}) & 0 & c_1(\boldsymbol{k}) \\ c_2(\boldsymbol{k}) & c_1(\boldsymbol{k}) & 0 \end{bmatrix}, \tag{2}$$

where $\epsilon$ and $t$ are $k$-independent parameters, and $c_i(\boldsymbol{k}) = \cos\left(\frac{1}{2}\boldsymbol{k}.\boldsymbol{a}_i\right)$. The eigenvalues follow from the secular equation,

$$det[H(\boldsymbol{k}) - (\epsilon + 2t\lambda)I_3] = 0 \tag{3}$$

which gives $\lambda_0 = -1$ and $\lambda_\pm(c_i(\boldsymbol{k})) = \frac{1}{2}\left(1 \pm \sqrt{4c_i(\boldsymbol{k})^2 - 3}\right)$, respectively. The eigenvalue $\lambda_0$ corresponds to a flat band as $\lambda_0$ does not dependent on $c_i(\boldsymbol{k})$ and hence not on $\boldsymbol{k}$, see Figure S1(b).

We now consider a decorated kagome lattice by adding another atom, centered in only one of the triangles of the kagome lattice. The coordinates of this fourth atom are given by,

$$\boldsymbol{r}_4 = \frac{1}{3}(\boldsymbol{a}_1 + \boldsymbol{a}_2). \tag{4}$$

The tight-binding Hamiltonian now reads,

$$H(\boldsymbol{k}) = \epsilon I_4 + 2t \begin{bmatrix} 0 & c_3(\boldsymbol{k}) & c_2(\boldsymbol{k}) & \delta_1(\boldsymbol{k}) \\ c_3(\boldsymbol{k}) & 0 & c_1(\boldsymbol{k}) & \delta_2(\boldsymbol{k}) \\ c_2(\boldsymbol{k}) & c_1(\boldsymbol{k}) & 0 & \delta_3(\boldsymbol{k}) \\ \delta_1^*(\boldsymbol{k}) & \delta_2^*(\boldsymbol{k}) & \delta_3^*(\boldsymbol{k}) & 0 \end{bmatrix}, \tag{5}$$

where $\delta_j(\boldsymbol{k}) = \frac{t'}{2t} e^{ik.(r_4 - r_j)}$ and $s_i = \sin\left(\frac{1}{2}\boldsymbol{k}.\boldsymbol{a}_i\right)$. Interestingly, the characteristic flat band of the kagome lattice is unaffected by this additional atom, as

$$2i[\delta_1^* s_1 - \delta_2^* s_2 + \delta_3^* s_3] = e^{-ik.\left(-\frac{1}{6}\boldsymbol{a}_1 - \frac{1}{3}\boldsymbol{a}_2\right)}\left(e^{ik.\frac{1}{2}\boldsymbol{a}_1} - e^{-ik.\frac{1}{2}\boldsymbol{a}_1}\right) -$$
$$e^{-ik.\left(\frac{1}{3}\boldsymbol{a}_1 - \frac{1}{6}\boldsymbol{a}_2\right)}\left(e^{ik.\frac{1}{2}\boldsymbol{a}_2} - e^{-ik.\frac{1}{2}\boldsymbol{a}_2}\right) + e^{-ik.\left(-\frac{1}{6}\boldsymbol{a}_1 - \frac{1}{6}\boldsymbol{a}_2\right)}\left(e^{ik.\left(-\frac{1}{2}\boldsymbol{a}_1 + \frac{1}{2}\boldsymbol{a}_2\right)} - e^{-ik.\left(-\frac{1}{2}\boldsymbol{a}_1 + \frac{1}{2}\boldsymbol{a}_2\right)}\right) = 0$$

(6)